# On the Impact of Non-Linear Phase-Noise on the Assessment of Long-Haul Uncompensated Coherent Systems Performance

Y. Jiang[(1)], A. Carena[(1)], P. Poggiolini[(1)], F. Forghieri[(2)]

[(1)] Politecnico di Torino, DET, corso Duca degli Abruzzi, 24,10129 Torino, Italy, poggiolini@polito.it
[(2)] Cisco Photonics Italy srl, via Philips 12, 20900 Monza, fforghie@cisco.com

**Abstract** *We accurately characterize nonlinear phase noise in uncompensated coherent optical systems. We find that, though present, its impact on system performance is typically negligible in a wide range of practical system scenarios.*

## Introduction

Various non-linear propagation models, among which the GN-model [1],[2], resort to the simplifying approximation that non-linearity-induced noise (or NLI, non-linear interference), in uncompensated transmission systems, approximately behaves as complex Gaussian-additive (GA) noise, statistically independent of either ASE noise or the signal itself. This assumption is not exact, but the good predictive power achieved using it [2]-[5], seemed to suggest that it is accurate-enough for system performance assessment.

Recently, however, it has been pointed out that in certain scenarios, such as with ideally lossless fiber or with very short spans, and in single-polarization, a very large non-additive phase-noise (PN) component of NLI may be generated [6],[7]. In addition, it has been highlighted that such PN has a long time-correlation, up to tens or even hundreds of symbols [6],[8]. Based on these results, the overall possibility of NLI being approximated as GA noise has been challenged.

In this paper we investigate this important issue by carrying out an in-depth simulative span-by-span characterization of PN in UT links using realistic span lengths and amplification schemes, in dual polarization. We find that nonlinear PN is indeed present, but its strength is small. This suggests that in practical system scenarios the GA approximation is adequate for carrying out system performance assessment. We further test this finding in a realistic PM-QPSK system scenario where we show that the assumption of Gaussian-additive NLI noise leads to a system maximum reach prediction that is accurate to within 0.1 dB of simulations.

## Simulation set-up and post-processing

This simulative study concentrates on phase noise in WDM PM-QPSK transmission. The transmitter outputs 9 channels operating at 32 GBaud, with raised-cosine spectra (roll-off 0.05). The channel spacing is 33.6 GHz so that transmission is quasi-Nyquist. We analyze links using two fiber types, SMF: $\alpha$ =0.22 dB/km, $\beta_2$ =20.8 ps$^2$/nm, $\gamma$ =1.3 1/(W·km); NZDSF: $\alpha$ =0.22 dB/km, $\beta_2$ = 4.7 ps$^2$/nm, $\gamma$ =1.5 1/(W·km). The span length is either 100 or 60km. All simulations are dual-polarization. Lumped amplification is assumed but no ASE noise is injected (similarly to [6],[7]) to avoid masking NLI noise.

We characterize phase noise at the receiver (Rx) as follows. The signal is demodulated using a matched filter. Polarization is recovered statically. No dynamic equalizer is present which could perturb the constellation or otherwise remove long-correlated noise components.

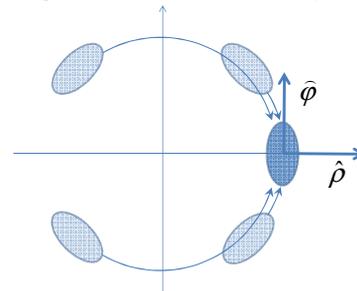

**Fig. 1:** Phase noise detection post-processing for PM-QPSK

Similar to [6],[7], we eliminate all single-channel (SC) non-linearity. We do it by subtracting the single-channel simulated signal from the WDM simulation, a procedure that yields very accurate results (see [9]). One reason for this is for easier comparison with [6],[7]. Another reason is that this provides a better picture of what could be the behavior of a fully loaded system, where SC effects would be relatively small. At any rate, we provide indications on SC effects, as well.

The signal is analyzed as shown in Fig.1. The four constellation points are analytically rotated so that they superimpose on the horizontal axis, in such a way that the major and minor axes of the four ellipses shown in figure are aligned. PN, if present, shows up on the tangential $\hat{\varphi}$ axis.

To study it in detail, we estimated the auto- and cross-correlation functions of the $\hat{\varphi}$ and $\hat{\rho}$ NLI components, that is: $R_{\hat{\varphi}\hat{\varphi}}(k)$, $R_{\hat{\rho}\hat{\rho}}(k)$ and $R_{\hat{\varphi}\hat{\rho}}(k)$ where $k$ is the delay in number of symbol times.

The variances of the $\bar{\varphi}$ and $\hat{\rho}$ NLI components are given by $\sigma_{\hat{\varphi}}^2 = R_{\hat{\varphi}\hat{\varphi}}(0)$, $\sigma_{\hat{\rho}}^2 = R_{\hat{\rho}\hat{\rho}}(0)$. If substantial PN was present, we would expect $\sigma_{\hat{\varphi}}^2 \gg \sigma_{\hat{\rho}}^2$, i.e., elliptic noisy signal "points" would show up as pictured in Fig.1, similar to results shown in [6],[7]. Also, if PN had a long-correlated component, it would show up as non-zero values of the $R_{\hat{\varphi}\hat{\varphi}}(k)$ curve for $|k| \neq 0$.

**Phase noise characterization results**

In Fig.2 we plot the measured results on the center channel, vs. number of spans $N_{span}$. We show results on one polarization, the other being identical. All results are normalized vs. $P_{ch}^3$, where $P_{ch}$ is the power per channel. According to theoretical models, such normalization makes the results launch-power independent [1],[2],[6],[7]. The simulations shown in Fig.2 used $P_{ch} = -3\text{dBm}$. We verified that identical plots are found at $P_{ch}$ both -6 and -1dBm.

Fig.2 shows that after one span $\sigma_{\hat{\varphi}}^2$ is about 2.5 dB larger than $\sigma_{\hat{\rho}}^2$, both for 60 and 100km NZDSF spans. However, already at ten spans the difference is down to 0.8 and 0.5dB (respectively) and tends to vanish at higher span count. Similar variance means that approximately "circular" noise shows up in the signals post-processed as in Fig.1 (small ellipticity of the constellation points). To provide visual evidence, in Fig.3 we show a plot structured as Fig.1, extracted from the simulation at 20 spans of NZDSF (100km spans), where $\sigma_{\hat{\varphi}}^2 / \sigma_{\hat{\rho}}^2 = 0.37\text{dB}$. The individual signal points, as well as the "superimposition" one, show in fact quite modest "ellipticity".

Regarding the presence of a long-correlated PN component, in Fig.4 we show the PN autocorrelation function $R_{\hat{\varphi}\hat{\varphi}}(k)$ for NZDSF (100km spans) at 20 spans, which presents a sharp peak $R_{\hat{\varphi}\hat{\varphi}}(0)$ (i.e., the variance $\sigma_{\hat{\varphi}}^2$) and then a slowly decaying trend vs. the delay $k$. Both [6],[8], predicted a slowly-decaying $\bar{\varphi}$ autocorrelation, so our results agree with this prediction. However, such long-correlated component appears to be small. A possible measure of its strength is the ratio $R_{\hat{\varphi}\hat{\varphi}}(1) / \sigma_{\hat{\varphi}}^2$ where $R_{\hat{\varphi}\hat{\varphi}}(1)$ is the correlation of PN at one symbol delay. Both Fig. 2 and Fig.4 show that such ratio is small and Fig.2 shows that it steadily goes down along the link. At 20 spans, it is -10 and -8 dB, for 100 and 60 km spans, respectively. Finally, in Fig.2 we also plot the cross-correlation function $R_{\hat{\varphi}\hat{\rho}}(k)$ for $k = 0$ and 1. These curves are less than -20dB vs. $\sigma_{\hat{\varphi}}^2$ and $\sigma_{\hat{\rho}}^2$. Not shown, $R_{\hat{\varphi}\hat{\rho}}(k)$ stays this low for any value of $k$. These results indicate that the $\bar{\varphi}$ and $\hat{\rho}$ NLI components are essentially uncorrelated.

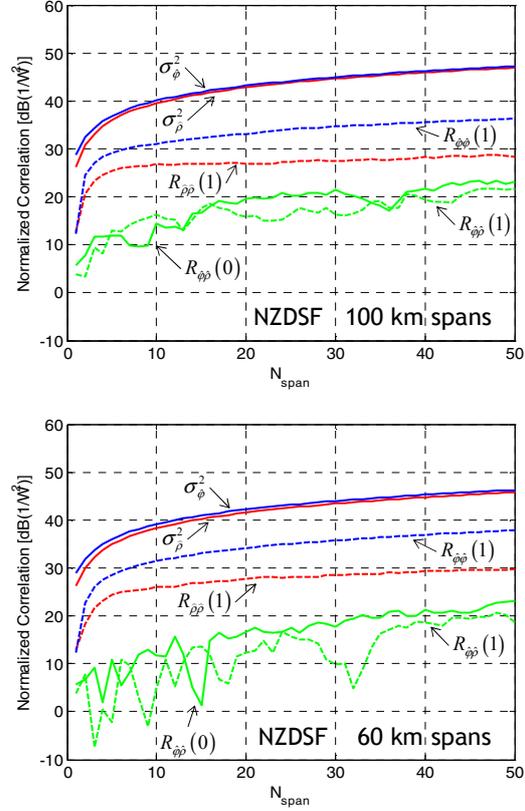

**Fig. 2:** Plots of variances and auto- and cross-correlation functions of the $\bar{\varphi}$ and $\hat{\rho}$ components of NLI noise, vs. number of spans (single-channel non-linearity removed). Correlation arguments are in symbol-time intervals. The system is 9-channel PM-QPSK, 32GBaud, quasi-Nyquist.

When single-channel non-linearity is not suppressed, at 10 spans of NZDSF, the ratio $\sigma_{\hat{\varphi}}^2 / \sigma_{\hat{\rho}}^2$ goes up slightly, to 1 and 0.7 dB (60 and 100km spans, resp.), still tending to vanish vs. $N_{span}$. Significantly, the $\sigma_{\hat{\varphi}}^2 / \sigma_{\hat{\rho}}^2$ ratio values for SMF (not shown for lack of space) always show *less* phase-noise than over NZDSF.

**Test-case of system performance evaluation**

For the 100km-spans NZDSF system used for Fig.2, we predicted the maximum reach using the non-linear OSNR, which strongly relies on the Gaussian-additive NLI assumption:

$$\text{OSNR}_{NL} = \frac{P_{ch}}{P_{ASE} + P_{NLI}} \quad (1)$$

$P_{ASE}$ was calculated assuming an EDFA noise figure of 5dB. For $P_{NLI}$, we used the total NLI variance found through simulation (including single-channel effects). For $P_{ch}$ we used the actual useful signal power, measured on the constellation by replacing each constellation point with its local average. We assumed a target BER of $2 \cdot 10^{-3}$. The results are shown in Fig.5 as a thin red line. The red squares are instead found by direct Monte-Carlo error count, over 132,000 simulated symbols. For details on the split-step simulation technique used, see [2].

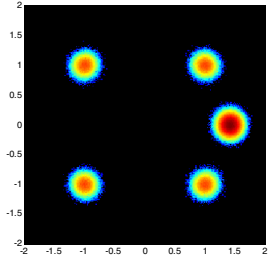

**Fig. 3:** Center channel constellation and "superimposed" signal point, structured as Fig.1, for a simulated case: 20 spans of NZDSF, span length 100km, 9-channel PM-QPSK, 32GBaud, quasi-Nyquist. Single-channel NLI removed.

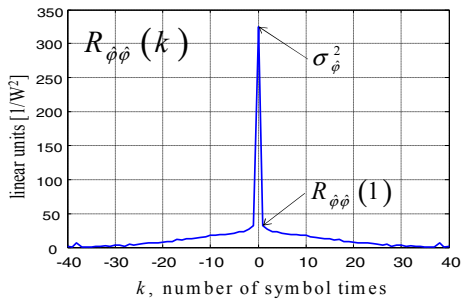

**Fig. 4:** Plot of the normalized autocorrelation function of the $\hat{\varphi}$ component of NLI noise, $R_{\hat{\varphi}\hat{\varphi}}(k)$, vs. number of symbol intervals, for NZDSF at 20 spans, span length 100 km. Single-channel NLI removed.

If PN was significant, or in general NLI noise was markedly non-GA and perhaps substantially dependent on signal and ASE, then the maximum reach found using the GA assumption Eq.(1) would noticeably differ from direct Monte-Carlo error-count results. Instead, the very good match of the two approaches suggests that the GA assumption and (1) are adequate for system performance assessment, in agreement with the results of the previous section. The results over SMF (not shown for lack of space) present similar very good agreement. Fig.5 also shows (dashed black line) the prediction found using a recently proposed improved-accuracy variant of the GN model (the EGN-model [9], based on an extension and generalization of [6]). The EGN model was used to find $P_{\text{NLI}}$ which was then placed into Eq. (1). Here too a very good match with simulation results is found, further upholding the adequacy of the GA assumption.

**Discussion and conclusion**
Our simulative results present a detailed investigation of phase-noise within NLI. We found it to be present, but its strength is limited and has little impact on system performance for typical PM-QPSK EDFA-amplified systems, even assuming low-dispersion fiber and short spans. Not shown for lack of space, we found similar results for PM-16QAM.

Our findings are also valid for those hybrid Raman/EDFA configurations (with backward-pumped Raman) where the span output power is at least 6 dB lower than the span input power.

In conclusion, the approximation consisting of considering non-linearity-generated noise as additive and Gaussian appears to be adequate for system performance prediction in a wide range of realistic system configurations. Work is in progress to address all-Raman systems.

This work was supported by CISCO Systems within a SRA contract

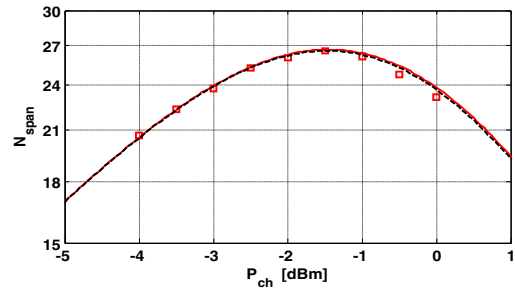

**Fig. 5:** 9-channel PM-QPSK, 100km-span NZDSF. Squares: Monte Carlo error count. Thin red line: Eq. (1) with NLI power measured on simulations. Dashed black line: Eq. (1) with NLI power estimated through the EGN model [9].